\documentclass[runningheads, anonymous]{llncs}
\usepackage[T1]{fontenc}
\usepackage{graphicx}
\usepackage[numbers]{natbib}
\usepackage{mathtools}
\usepackage[linesnumbered,ruled,vlined]{algorithm2e}
\usepackage{booktabs} 
\usepackage{multirow}
\usepackage{xcolor}
\usepackage{amsmath}
\usepackage{amssymb}
\usepackage[inline]{enumitem}
\usepackage[font=small,labelfont=bf]{caption}

\SetKwInOut{Input}{Input}
\SetKwInOut{Output}{Output}

\DeclarePairedDelimiter{\ceil}{\lceil}{\rceil}

\definecolor{mygray}{rgb}{0.75, 0.75, 0.75}

\begin{document}




\title{FedFNN: Faster Training Convergence Through Update Predictions in Federated Recommender Systems}

\author{
Francesco Fabbri\inst{1} \and
Xianghang Liu\inst{2} \and
Jack R. McKenzie\inst{2} \and
Bartlomiej Twardowski\inst{3} \and
Tri Kurniawan Wijaya\inst{3}
}

\titlerunning{FedFNN: Faster Training Convergence in Federated Systems}
\authorrunning{F. Fabbri, X. Lui, J. McKenzie, B. Twardowski, T.K. Wijaya }

\institute{
\inst{1} Eurecat, Barcelona, Spain \\
\email{francesco.fabbri@eurecat.org}
\and
\inst{2} Huawei UK Research Centre, London, UK \\
\email{xianghang.lui@huawei.com} \\\email{jack.mckenzie1@huawei.com}
\and
\inst{3} Huawei Ireland Research Centre, Dublin, Ireland \\
\email{bartlomiej.twardowski@huawei.com} \\\email{tri.kurniawan.wijaya@huawei.com}
}

\titlerunning
\authorrunning
\maketitle

\begin{abstract}
Federated Learning (FL) has emerged as a key approach for distributed machine learning, enhancing online personalization while ensuring user data privacy. Instead of sending private data to a central server as in traditional approaches, FL decentralizes computations: devices train locally and share updates with a global server. A primary challenge in this setting is achieving fast and accurate model training—vital for recommendation systems where delays can compromise user engagement. This paper introduces FedFNN, an algorithm that accelerates decentralized model training. In FL, only a subset of users are involved in each training epoch. FedFNN employs supervised learning to predict weight updates from unsampled users, using updates from the sampled set. Our evaluations, using real and synthetic data, show: $(i)$ FedFNN achieves training speeds 5x faster than leading methods, maintaining or improving accuracy; $(ii)$ the algorithm's performance is consistent regardless of client cluster variations; $(iii)$ FedFNN outperforms other methods in scenarios with limited client availability, converging more quickly.
\end{abstract}



\section{Introduction}

\emph{Federated Learning} (FL) has become a pivotal formulation for distributing the computation of machine learning (ML) models. This approach improves online services, such as content personalization, while simultaneously preserving user privacy. This paradigm is gaining significant traction, particularly for use cases like next-word prediction or item suggestions. This is because a personalized output can be generated on-device (e.g., smartphone, tablet), negating the need to transfer sensitive data to an external server \cite{flanagan2020federated, mcmahan2017communication, ye2020edgefed, zhang2021subgraph}.

Unlike traditional centralized learning approaches where private user data is sent to a central server for synchronous training, an FL setting distributes computations across multiple devices. A server coordinates the local training of devices and aggregates their updates. This structure mitigates privacy risks like personal data leakage, misuse, or abuse since client training data remains undisclosed to the central server.

In the FL framework, mobile devices (\textbf{clients}) train their ML model locally, sharing only model updates—not the raw historical data—with a centralized unique aggregator (\textbf{server}) \cite{fedfast, kairouz2019advances, mcmahan2017communication, li2020federated}. The server coordinates training and distributes model updates to clients at each round.

In this work, our focus is on applying FL to recommender systems. These systems are trained on user history data and predict users' preferences for items. Both user and item data are typically represented through a low-dimensional embedding, a critical component for most state-of-the-art recommendation models \cite{ricci2011introduction, zhang2019deep}.

The training mechanism of FL models is delineated in Figure 1a. Initially, a subset of clients is chosen. The client's historical interactions (denoted by 1b - Client History) are then transformed into embeddings (indicated as 2a - Embeddings). These embeddings undergo local training (2b - Local Training) and are subsequently updated. Post this local training phase, the embedding alterations are transmitted to the server (3a - Server). The server aggregates these updates (3b - Model Aggregated) and relays the new data to all clients, enabling them to generate recommendations (4 - Recommendations).
\begin{figure*}[t]
\begin{tabular}{cc}
    \includegraphics[width=0.50\linewidth]{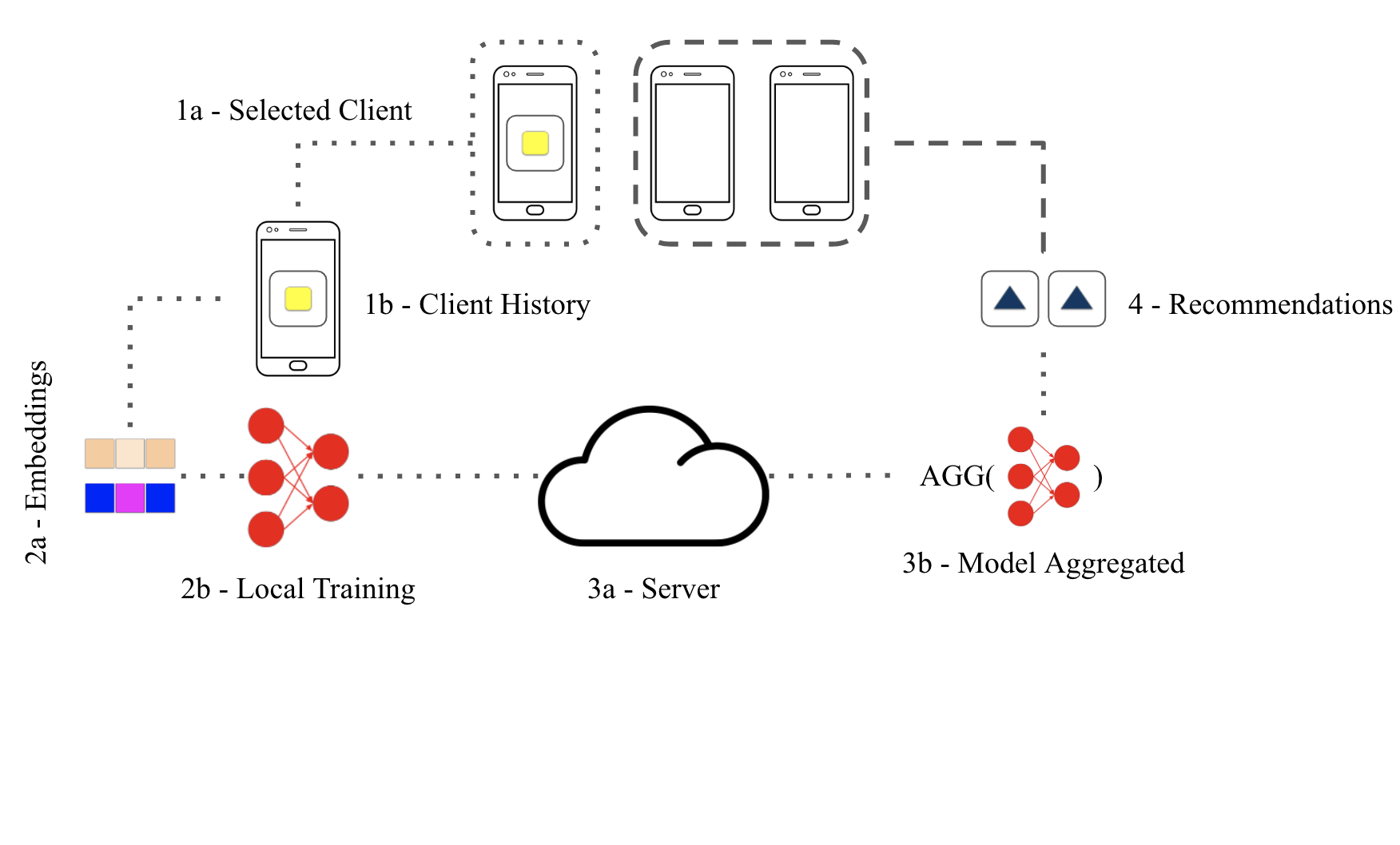}  
    &
    \includegraphics[width=0.50\linewidth]{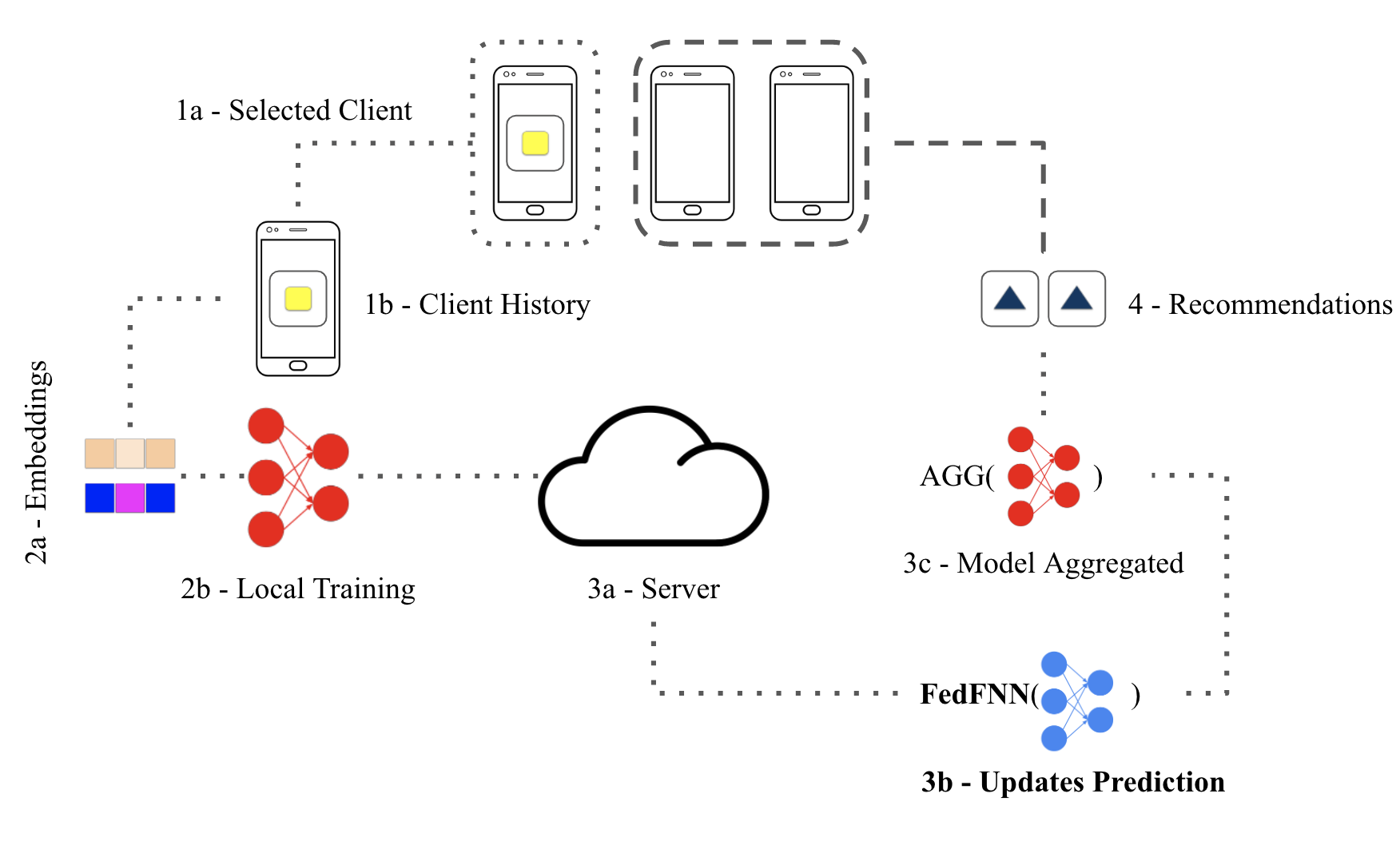}\\
    (a) FedAvg & (b) FedFNN
 \end{tabular}
\caption{On the upper figure (a), an illustration of the Federated Learning training process: $(1)$ the interactions of the client with the device are converted to vector embeddings (for both items and users); $(2)$ then the local training on the selected devices; $(3)$ the new training information are then sent to the server to be aggregated. $(4)$ the server sends the updates to all the clients which can generate the recommendations. On (b) the introduction of FedFNN, which generates the predictions of the unselected clients before sending the new updates.}
    \label{fig:idea}
\end{figure*}

This pipeline, proposed by \cite{mcmahan2017communication} represents the standard for training FL models in which, the representations for the trained users and items will be aggregated as an average on the server, before then being redistributed among untrained clients. 

FL training can occasionally be hindered by specific data characteristics, leading to protracted training durations and sub-optimal model performance. Notable among these are: $(i)$ \emph{sparsity} - a scenario where users typically engage with only a limited array of items; $(ii)$ \emph{heterogeneity} - wherein each client possesses data pertaining exclusively to its user; and $(iii)$ \emph{poor client availability} - during each training round, only a subset of clients (referred to as \textbf{delegates}) participate, leaving a significant majority (\textbf{subordinates}) inactive.

To circumvent these challenges and expedite training, we introduce \texttt{FedFNN}. This innovative algorithm harnesses a meta-network to learn and predict the embedding updates from local client training. This meta-network can proactively update embeddings even for clients that weren't chosen in the initial selection, thus accelerating convergence, as visualized in Figure 1b.

Our model, \texttt{FedFNN}, amalgamates recent advancements targeting training speed optimization \cite{fedfast}. It has demonstrated superiority in both convergence speed and prediction accuracy. The model utilizes a supervised learning approach to anticipate updates for unseen clients during each training iteration. Its efficacy has been empirically validated on both synthetic and real-world data sets. Our evaluations pitted \texttt{FedFNN} against the competitive \texttt{FedFast} algorithm and various other baselines. Results showcased \texttt{FedFNN}'s prowess, as it consistently delivered results at least five times faster in terms of the number of iterations required to achieve comparable accuracy levels.

To simulate scenarios of low client availability and data sparsity, we employed a synthetic data generator. This tool facilitates the artificial generation of user-item interaction data, allowing the creation of distinct client clusters based on preference. Our findings underscore the robustness of \texttt{FedFNN}, both in terms of convergence speed and prediction accuracy, regardless of the client group size or availability.

Additionally, we conducted an ablation study to decipher the contribution of various \texttt{FedFNN} components. Specifically, we delved into the realms of item update aggregation and subordinate predictions. Our investigations revealed that our method for predicting subordinates emerged as the most optimal, regardless of the item update strategy employed.

\noindent\textbf{Paper Structure.} The subsequent sections of this paper are organized as follows: Section~\ref{sec:rw} delves into relevant literature. In Section~\ref{sec:model}, we elucidate the background and provide a formal description of our algorithm. Section~\ref{sec:exp} details our experimental design and presents the results. Concluding remarks and potential avenues for future research are discussed in Section~\ref{sec:disc}


\section{Related Work}
\label{sec:rw}


Federated Learning has emerged recently as popular research area, since it allows exploitation of the vast amount of data collected on personal devices through privacy-preserving technologies \cite{kairouz2019advances, li2020federated}. 
\cite{mcmahan2017communication} was the first to introduce the definition of Federated Learning, 
with the Federated Averaging (\texttt{FedAvg}) algorithm that is capable of increasing the training speed of classification models through the sharing of weight updates among sampled clients \cite{mcmahan2017communication}. 
%

Parallel to the \texttt{FedAvg} improvements, there is a new line of work proposing increased accuracy of models in the presence of heterogeneous data distributions among devices (i.e. an imbalanced distribution of samples among the different devices), also called \emph{statistical heterogeneity} \cite{li2018federated, wang2020tackling, karimireddy2020scaffold}. \cite{li2018federated} propose \texttt{FedProx}, which modifies \texttt{FedAvg} algorithm by including a proximal term, reducing differences between local and global updates \cite{li2018federated}. 




The FL paradigm has been mainly investigated for supervised learning tasks (e.g. binary and multi-label classification). Due to its effectiveness of generating on-device personalised recommendations without violating the user privacy, it has also been applied to the area of information retrieval and recommendation algorithms \cite{wu2021fedgnn, fedfast, anelli2021federank, kalloori2021horizontal, liang2021fedrec++}.

Most of the contributions focus on the matrix factorisation algorithm, which represents the standard to capture latent dimensions of user-item interactions \cite{he2017neural, koren2009matrix}.
\cite{anelli2021federank} introduce \texttt{FedeRank}, a federated recommendation algorithm  which allows users to control the portion of data they want to share \cite{anelli2021federank}; 
it focuses on the accessibility of personal data, which is an aspect not covered by \texttt{FedFNN}. 
\cite{lin2020fedrec} propose \texttt{FedRec}, a federated model for predicting explicit feedback \cite{lin2020fedrec}; in contrast, our work focuses on implicit feedback. 
Additionally, their objective is to generate performances as close as possible to the centralised version of the model, whereas our focus is to improve training speed. 
%

\cite{fedfast} propose \texttt{FedFast} algorithm that proposes to
to speed up the recommender system training by clustering the users \cite{fedfast}. 
Specifically, they propose: 
$(i)$ a model to update unsampled clients, 
$(ii)$ an item update strategy which looks at client contribution weighting and 
$(iii)$ a sampling technique that is coherent with the clustering approach. 
Compared to our work, they rely on the assumption of having clustered embeddings in both, the input data distribution and the clients sampled for the local update. We show through an ablation study how their clustering approach does not significantly help in the prediction task, presenting performances close to the baseline (with no prediction for the unsampled clients).
Our work aims to capture similarities at individual and not group level, between user embeddings. 
We extensively compare \texttt{FedFNN} to \texttt{FedFast} in Section~\ref{sec:exp}, 
showing the effectiveness of our solution. 

\texttt{FedFNN} intuition is built on the idea of predicting differences in the next weights updates of the subordinate clients, which intuitively can be connected to \cite{He_2016_CVPR}, a popular architecture achieving state-of-the-art performances in Computer Vision tasks, where the hidden layers are enhanced with the prediction of the difference between the new and old weights. \cite{ha2016hypernetworks} proposes hypernetworks, strongly related to our work, where a single network is used to generate the weights of another architecture \cite{ha2016hypernetworks, zhang2018graph}. 
In hypernetworks, one architecture is trained and the other predicted, while in our case, \texttt{FedFNN} is trained to predict the updates of the weights of the main model, which then uses the updates for the local training. 
Another relevant line of research connected to hypernetworks is meta-learning, which has been also explored in connection with Federated Learning~\cite{lin2020meta,chen2018federated}. 
Specifically, in those approaches a meta recommender system is designed, able to generate item embeddings with a meta network.

\texttt{FedFNN} is built on Matrix Factorisation algorithm, which represents the standard for recommender systems in presence of implicit feedback \cite{koren2009matrix}. However recently, models more sophisticated than that have also been proposed. Among many, \cite{wu2020fedctr} introduced a federated framework for online advertising to 
predict CTRs of native ads from multi-platform user behavior \cite{wu2020fedctr}, 
while \cite{wu2021fedgnn} introduce a graph neural network architecture to predict high-order user interactions \cite{wu2021fedgnn}. \texttt{FedFNN} can be extended to other sophisticated models in a FL setting and we leave this task for the future work.

\section{Preliminaries}
\label{sec:model}

Federated Learning is characterised by a unique server which handles incoming messages from multiple clients. The server aggregates these messages to optimise a single global optimum. In particular the objective is to minimise:
\begin{equation}
f(v,w) = \text{min } \sum_{k = 1}^N  p_k F_k(v, w)
\end{equation}
where $N$ is the number of devices, 
$v$ the item embeddings, 
$w$ the user embeddings,
$F_k(v,w)$ the local objective for the device $k$, which measures the local empirical risk over data distribution $D_k$,
and the normalisation factor $p_k \in [0,1]$ applied to $F_k(v,w)$. 
Let the learnable parameters of the General Matrix Factorisation (\texttt{GMF}) algorithm be denoted by  $\{v, w, \phi \}$. The parameter set $\phi$ encompasses all the additional parameters intrinsic to the \texttt{GMF} algorithm, excluding $v$ and $w$ These parameters, although less emphasized in our context, are indispensable for the effective training of the \texttt{GMF}.

This model is widely used in designing recommendation algorithms and predicts user preferences through a low-dimension representation of the data. 

To learn the user preferences from the embeddings, it uses a fully connected layer that takes as input user and item embeddings and outputs the scores distribution for all the missing interactions \cite{he2017neural}.

This model can then be naturally extended in a federated setting using \texttt{FedAvg}, where a subset of all devices are selected for local updates. 
Algorithm \ref{algo:flalgo} presents the steps of \texttt{FedAvg};
each $k$-th sampled client (line 3) gets first a copy of the model to train on its own local data $D_k$, and subsequently the loss function $L$ optimised by the local model with respect to $\{v_k, w_k, \phi_k \}$ (lines 4,5):
\begin{equation}
\{v_k, w_k, \phi_k \} \leftarrow  \textbf{ A }(v, w, \phi, L, D_k),  \text{  for each } k \in S, 
\end{equation}
where $(v_k, w_k, \phi_k)$ are the learnable parameters updated by the delegate client $k$, \textbf{A} is the optimisation algorithm selected to estimate the parameters values and 
$S \subseteq \{1, \ldots, N\}$ is the indices of all the delegates. 
Since we work with implicit feedback data, a binary cross-entropy loss function is selected; for the optimisation algorithm the ADAM optimiser is chosen.

For each training round, the weights updated through the local models are aggregated to generate the global update (lines 6-9). In particular, among all the client embeddings included in the global model $w[D]$, when using \texttt{FedAvg}, only the embeddings of the delegates  $S$ (i.e. $w_k[k], k \in S$) will be updated: 
\begin{equation}
w^\textit{new}_k[k] \leftarrow w_k[k], \text{ for each } k \in S 
\end{equation}
In contrast, the item embeddings are aggregated in the server (lines 9-10) and then redistributed among all the clients.

\begin{algorithm}[tb]
  \caption{\textsc{FL Algorithm} }
  \label{algo:flalgo}
  \Input{set of item $I$, set of clients $U$, sample size $m$}
  Initialise: user embeddings $w^0$, item embeddings $v^0$\;
  \For{$t \gets 1,2,\ldots,T$}{
    $S_t \gets$ ClientSampling($U$, $m$)\;
	\For{client $k \in S_t$}{
	  $w_k^{t+1}, v_k^{t+1}  \gets$ LocalUpdate($w^t_k,v^t_k, k$)\;
    }
    $\Omega_t \gets w^{t+1}, S_t, m$\;
    \For{client $j \in U \backslash S_t$}{
	  $w_j^{t+1} \gets$ ClientPrediction($\Omega_t$)\;
    }
	$v^{t+1} \gets$ ItemUpdate$( v^t[S_t])$\;
  }
\end{algorithm}

Updating only a subset of clients represents a limitation that affects the number of rounds necessary to get good model performances. In this direction, subordinate clients updates are crucial to the convergence of the global model. 
In \texttt{FedAvg} the embeddings of the local clients are updated as follows:
\begin{equation}
w^{\textit{new}}[j] \leftarrow \frac{1}{|S|} \sum_{k \in S} w_k[k],  \text{  for each } j \in U \setminus S
\end{equation}

To update embeddings of subordinate clients, \texttt{FedFast} recently proposes a clustering-based model aggregation approach to update the embeddings of subordinate users with the averaged weight update of the selected users belonging to the same cluster \cite{fedfast}.
First, users are clustered based on their embeddings $C = \{c_i \mid i = 1, 2, \ldots\}$, 
where each cluster $c_i$ is a set of users. 
Then, the averaged weight update $\delta_{c_i}$ is computed for all clients in $c_i$:
\begin{equation}
\delta_{c_i} \leftarrow \frac{1}{|c_i \cap S|}\sum_{k \in c_i \cap S} w_k[k] - w[k], \text{ where  } i = 1, 2, \ldots, |C| 
\end{equation}

The new averaged update is used for modifying the weights of the subordinate users:  
\begin{equation}
w^{\textit{new}}[k] \leftarrow w[k] + \delta_{c_i}, \text{ for each } k \in c_i \setminus S 
\end{equation}


Despite being faster to converge than \texttt{FedAvg}, \texttt{FedFast} relies on two assumptions which may be difficult to maintain consistently: 
\emph{(i)} it assumes that clients can be clustered into distinct groups, which is not always the case (e.g. group of users presenting highly overlapping preferences); 
\emph{(ii)} even upon assuming the existence of the groups of preferences, fixing an \emph{a priori} number of clusters for the sampled clients relies on the feasibility to sample users from all the clusters at each round.
Moreover, since real data cannot be accessed in a unique pass, finding the correct number of clusters in the federated setting may require multiple tests before the final model training. 
Due to these limitations, as is found in our ablation study in Section~\ref{sec:exp}, 
their cluster-based update propagation produces little to no improvements over the training of the models.


\section{FedFNN}
To overcome the limitations mentioned above, we relax the assumption of the clustered preferences to the assumption that if two users are close in the embedding space, their embeddings would have similar updates after local learning. 
Under this new, relaxed assumption, we propose \texttt{FedFNN} that directly predicts the embedding weight updates for subordinates (i.e. users that are not directly sampled and updated in one epoch) using supervised regression models. 
This idea is also closely related to the class of meta-learning algorithms which directly generates the weights of the target model \cite{ha2016hypernetworks, zhang2018graph}.
However, instead of directly generating the embedding weights,
\texttt{FedFNN} predicts the embedding updates using those produced by the delegate clients. 

The training phase is performed on the data coming from the delegates to optimise a learning function $f(\cdot; \theta)$, using the previous user embeddings as input variable and the new embeddings updates as output:
\begin{equation}\label{eq:main}
\min_{\theta} : \mathbb{E}_{U} \left[ \ell\left( f\left( w[U]; \theta \right), w^{\textit{new}}[U] - w[U] \right) \right]
\end{equation}
The objective function is the expected RMSE taken over the distribution of all the clients $U$ as we are interested in the generalisation error of the update predictor. The architecture for $f(\cdot; \theta)$ proposed in the experiments with FedFNN is a multi-layer perceptron (MLP), which proves to be simple but effective with the data sets we use. 

Then, for each subordinate client we predict the updates
\begin{equation}
w^{\textit{new}}[j] \leftarrow e^{- \gamma t }  \ell(w[j]; \theta^*)  + (1 - e^{- \gamma t }) w[j] , \text{ for each } j \in U \setminus S
\end{equation}
where $\gamma$ is the discount factor and $t$ the number of running epochs. 
Both of these parameters are combined to generate an exponential decay and reduces the contribution of the new predictions after each iteration. 
The updates are predicted by minimising the error produced on average (RMSE) and since weights need to converge towards personalised values dependent on the client information, 
those predictions becoming less relevant in the long-run.

In contrast to \texttt{FedFast}, which uses unsupervised learning (e.g., k-means), \texttt{FedFNN} uses supervised learning that explicitly models the mapping from the client representation to the user embedding updates.
Additionally, \texttt{FedFNN} enables the introduction of more sophisticated machine learning models (e.g. Neural Networks), adding the ability to capture more information and increasing modeling power. 
Later, we show through an ablation study that updating the subordinate client embedding using \texttt{FedFNN} can generate significant improvements over the training of the models, which is not the case for the clustering method.
\begin{algorithm}[t]
  \caption{\textsc{FedFNN} }
  \label{algo:fedfnn}
  \Input{$\Omega_t \leftarrow\{ W_{t}, S, W_{t+1}[S]\}$, prediction function $g(.)$, number of delegates $m$, patience criteria $p$}
  \Output{$W_{t+1}[D]$, $RMSE_t$}
  \If{$t \leq p$ \textbf{or} $|1 - L(t)/L(t-p)| \geq \epsilon$}{
    $\delta[S] \leftarrow W_{t+1}[S] - W_t[S]$\;
    $\theta^*, RMSE_t \leftarrow \arg \text{ opt}_{\theta} \sum_{k \in S} \ell(g(w_t[k]; \theta), \delta[k])$\;
    \For{client $j \in D \backslash S_t$}{
      $w_{t+1}[j] \leftarrow e^{- \gamma t }  g(w_t[j]; \theta^*)  + (1 - e^{- \gamma t }) w_t[j]$\;
    }
  }
\end{algorithm}

\subsection{Optimisation} 

To find the optimal parameter $\theta$ in Eq.~\ref{eq:main}, we adopt a standard supervised learning approach based on cross validation. More specifically, for each communication round, we run 5-fold cross validation on the delegate clients, train $f(\cdot;\theta)$ on the training set using SGD based optimisation methods and approximate the objective in Eq.~\ref{eq:main} with the empirical loss on the validation set. During each round, we use grid-search to optimise the hyper-parameters on the validation set. Our set of hyper-parameters include: the network architectures, learning rates and dropout ratios for regularisation.
Moreover, since FedFNN is designed to optimise the training of the global model at early stage of the iterations, we use a \emph{patience} parameter $p$, similar to the early-stopping criteria used to avoid over fitting in supervised learning \cite{lutz1998early}, to stop predicting the subordinates embeddings. Specifically, at each iteration $t$ we monitor the loss function value of the global model with respect to the value at iteration $(t-p)$. Once the loss function stabilizes around a previous value, we stop using FedFNN. More precisely, we stop using FedFNN when $|1 - L(t)/L(t-p)| < \epsilon$ (where $\epsilon  = 0.01$), where $L$ represents the loss function of the global model. Algorithm \ref{algo:fedfnn} presents the whole process.


For the item updates, various  strategies can be applied to define the importance of the local update generated through the client $k$. Specifically, a coefficient $\alpha_k$ can be defined for each delegate client. In particular we implement \texttt{FedFNN} along with 3 different strategies:

\begin{itemize}
\item\textbf{W0}: The new local embedding updates are re-distributed through averaging the new changes among all the sampled items, where $\alpha_k = 1/m$, $\forall k \in S$

\item\textbf{W1}: The local item updates are distributed following the weighted average proposed in \cite{fedfast}, where the importance of weights updates in the current iteration is distributed according to the difference between the update and the embedding. 
Let 
\begin{equation}
Z_{t+1}[k] = \sum_{v_i \in D_k} ||v^i_{t+1}[k] - v^i_{t}||_1.
\end{equation}
Then, we set
\begin{equation}
\alpha_k = \frac{Z_{t+1}[k]}{\sum_{k \in S_t} Z_{t+1}[k]}.
\end{equation}
In this way, the clients presenting bigger updates are valued more.

\item\textbf{W2}: The local item updates are weighted by the number of samples included in each delegate client, where $\alpha_k = \frac{n_k}{\sum_k n_k}$, where $n_k$ is the number of interactions present in the client history.

\end{itemize}

\subsection{Sampling} 
In contrast to the \texttt{FedFast} algorithm, we do not add any constraint to the sampling strategy. We assume that in a practical, real use case scenario, sampling would be determined via external factors (e.g. client availability, connection quality, etc), not the algorithm being used.


\section{Experiments}
\label{sec:exp}

\begin{figure*}[t]

    \centering
    \includegraphics[width=.7\linewidth]{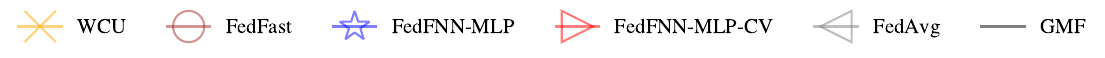}\\

    \begin{tabular}{cccc}
    \hspace{-3mm}\includegraphics[width=0.25\linewidth]{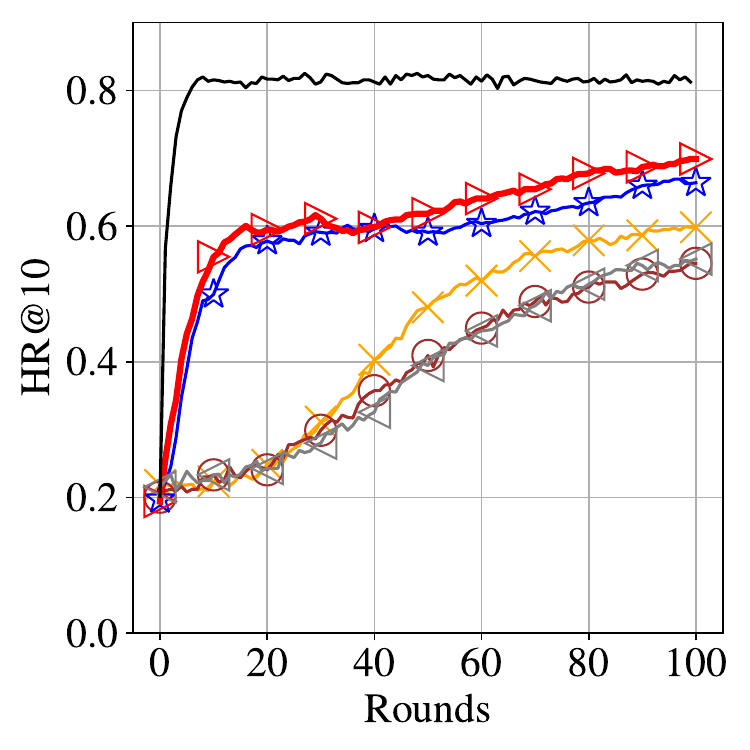} &
    \hspace{-3mm}\includegraphics[width=0.25\linewidth]{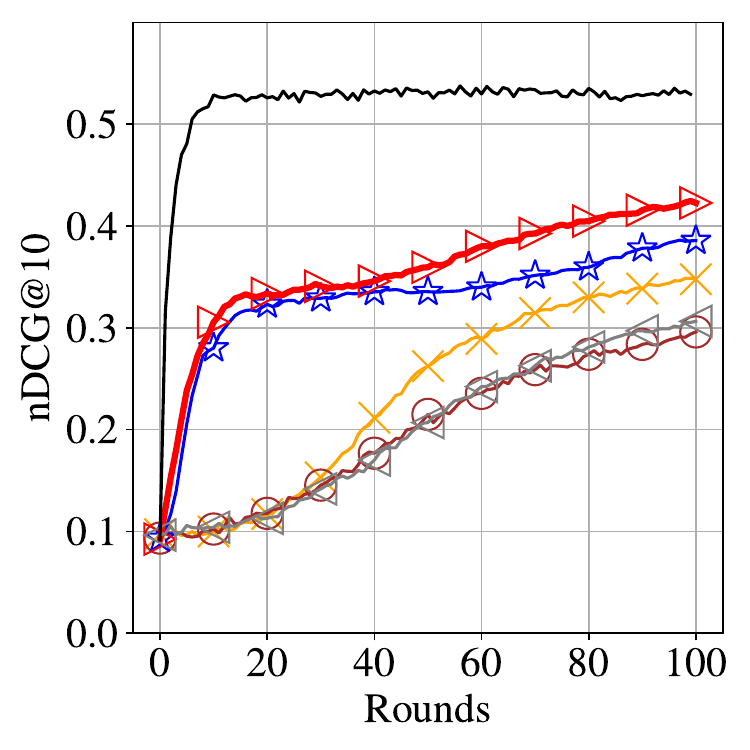}&
    \hspace{-3mm}\includegraphics[width=0.25\linewidth]{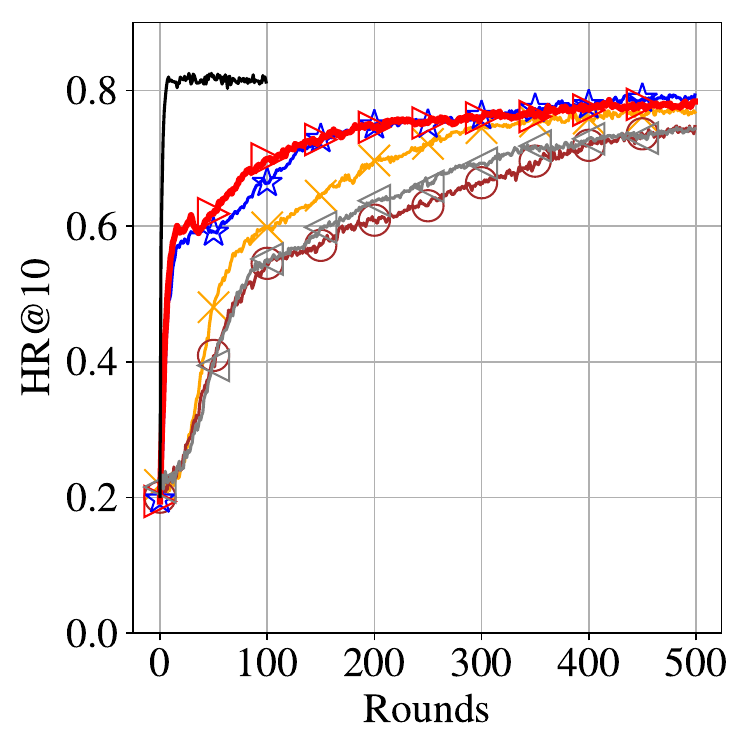} &
    \hspace{-3mm}\includegraphics[width=0.25\linewidth]{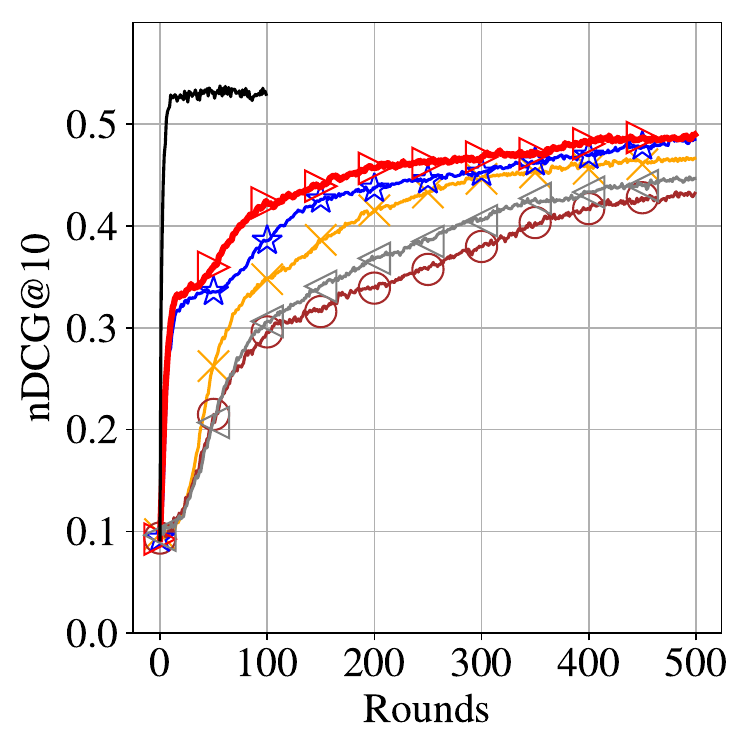}\\
    \end{tabular}\rotatebox[origin=c]{90}{(i)ML-100k}  
    \begin{tabular}{cccc}
    \hspace{-3mm}\includegraphics[width=0.25\linewidth]{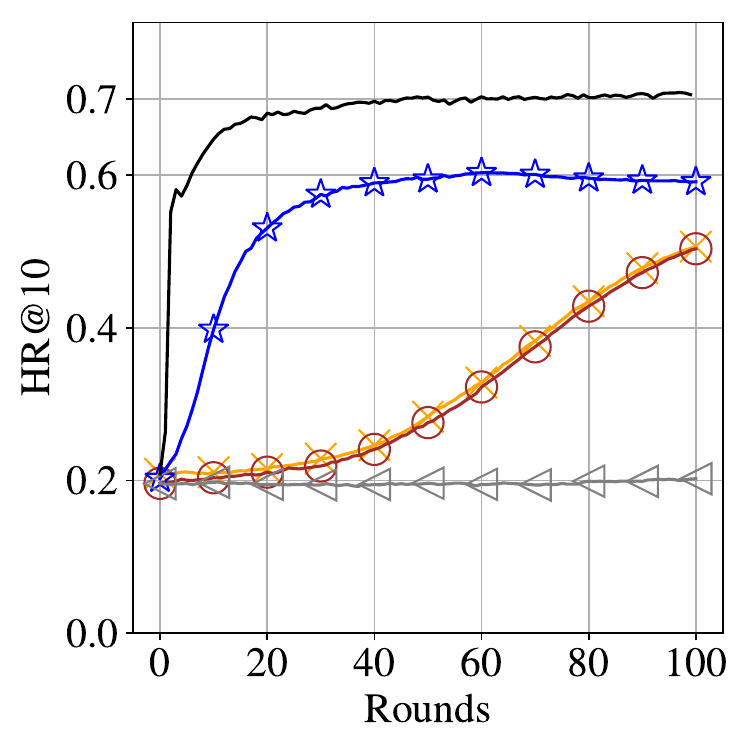} &
    \hspace{-3mm}\includegraphics[width=0.25\linewidth]{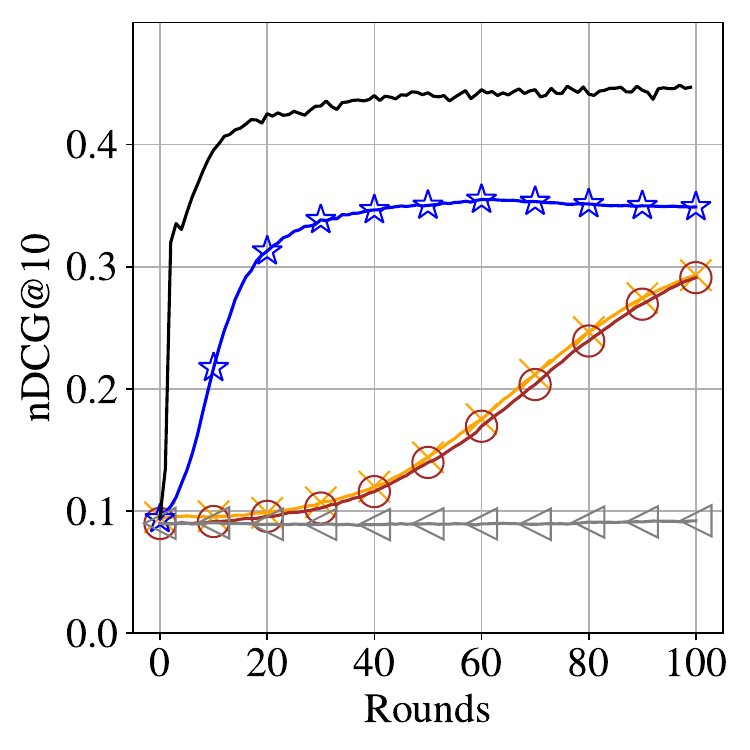}&
    \hspace{-3mm}\includegraphics[width=0.25\linewidth]{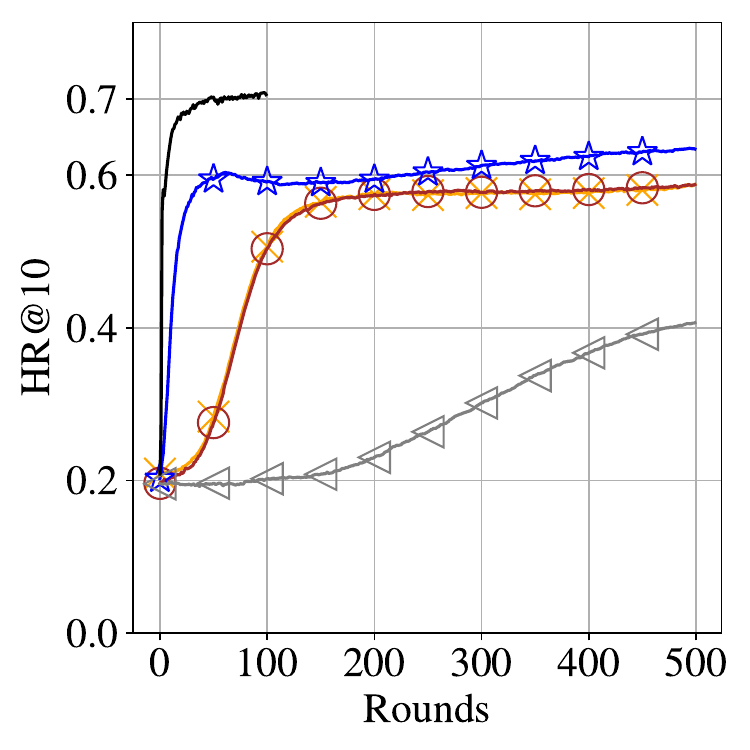} &
    \hspace{-3mm}\includegraphics[width=0.25\linewidth]{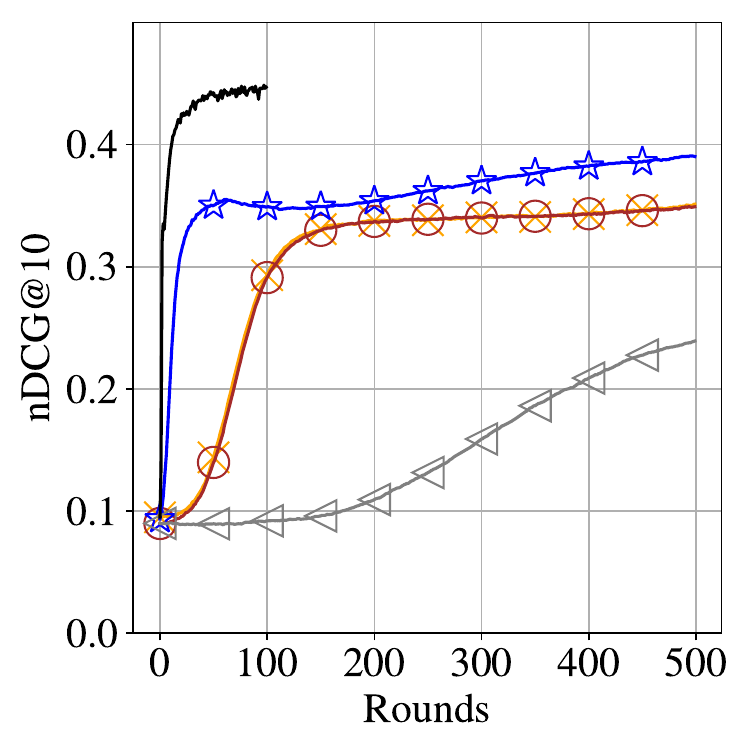}\\
    \end{tabular}
    \rotatebox[origin=c]{90}{(ii)YELP}\\ 
    \begin{tabular}{cccc}
    \hspace{-3mm}\includegraphics[width=0.25\linewidth]{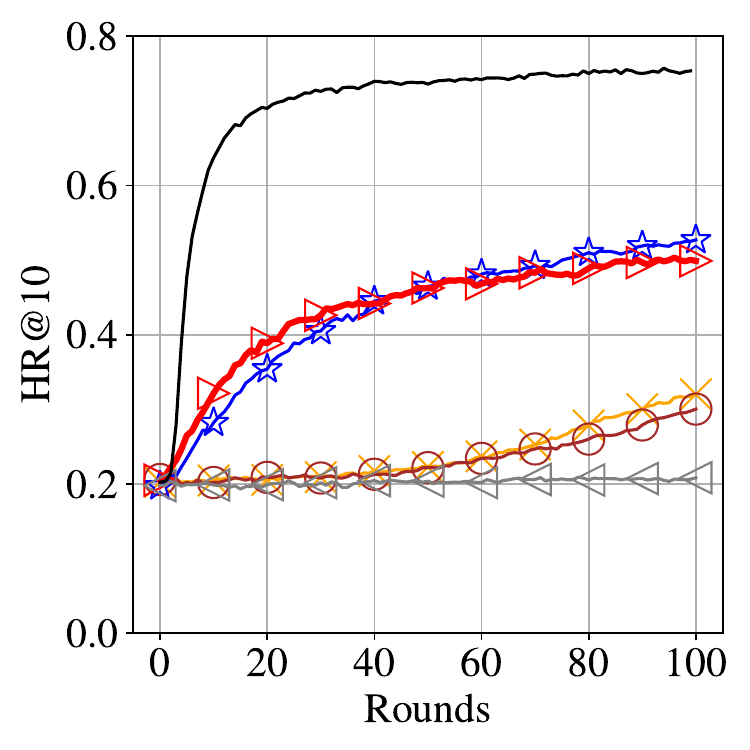} &
    \hspace{-3mm}\includegraphics[width=0.25\linewidth]{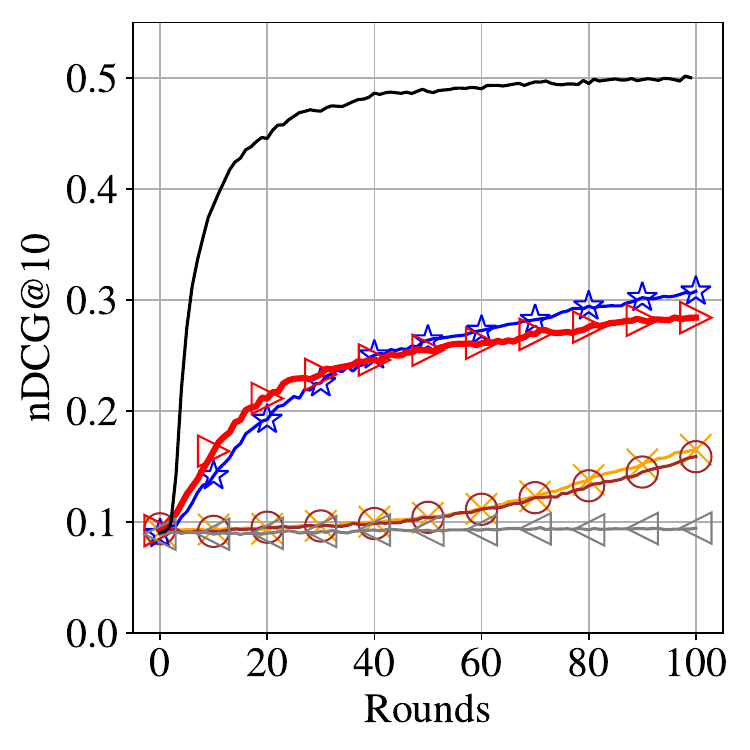}&
    \hspace{-3mm}\includegraphics[width=0.25\linewidth]{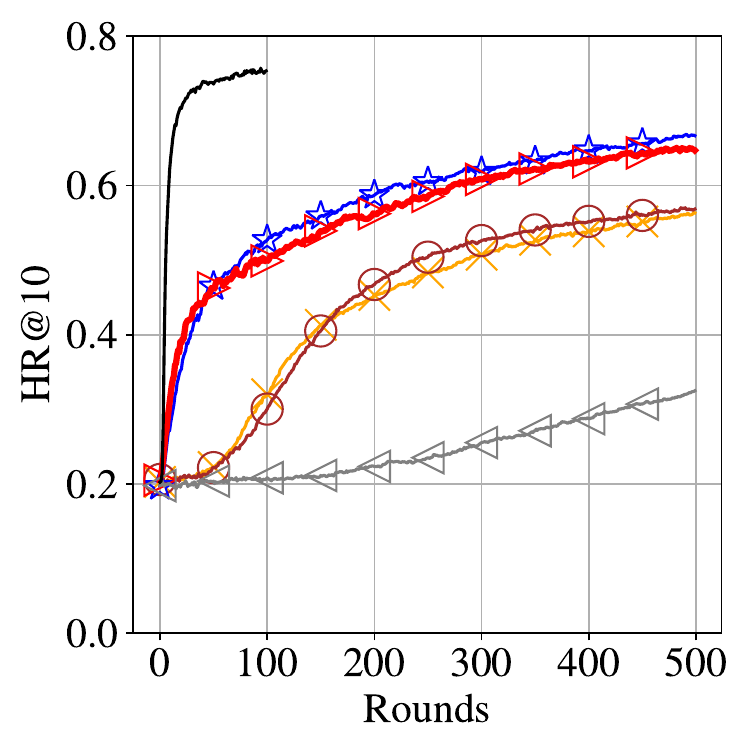} &
    \hspace{-3mm}\includegraphics[width=0.25\linewidth]{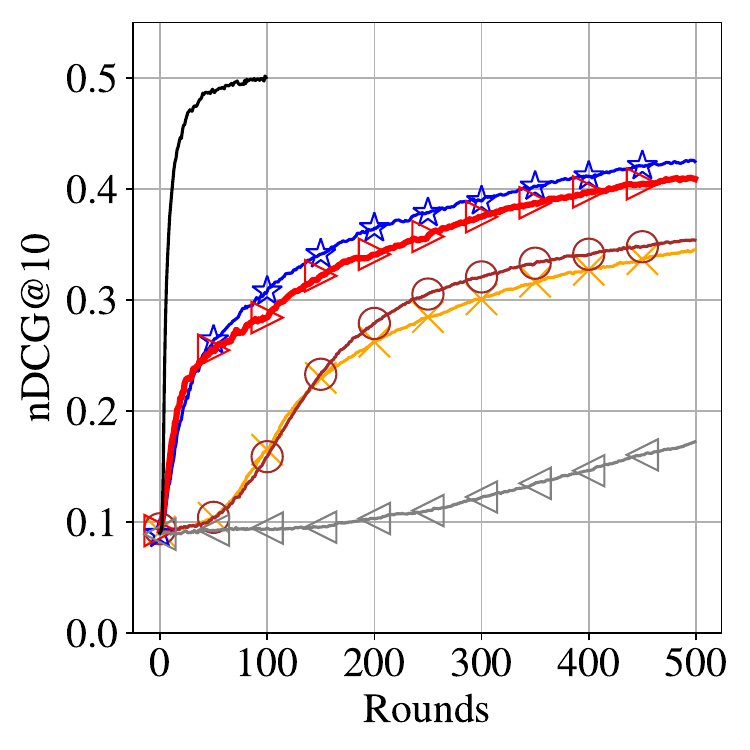}\\
    \end{tabular}
    \rotatebox[origin=c]{90}{(iii)AMAZON}

\caption{Comparison of FL models on three different data sets. The first two columns show only the first 100 iterations, while the third and fourth columns are showing up to 500 iterations.}
    \label{fig:comparison}
    \vspace{-3mm}
\end{figure*}

This section focuses on answering the following research questions: $(RQ1)$ What is the performance of \texttt{FedFNN} compared to the state-of-the-art in terms of accuracy and speed of convergence?; $(RQ2)$ Do the different strategies for updating both user and item embeddings significantly  increase the speed of convergence for FL models?; $(RQ3)$ How might distribution of user preferences affect the performances of \texttt{FedFNN}?; $(RQ4)$ How might client availability and sample size impact the performances of \texttt{FedFNN}?

\noindent \textbf{Data.} For the experimental section we use 3 different data sets: MovieLens, a popular movies reviews' collection used for bench-marking recommendation algorithms (\textbf{ML-100K}); Yelp, an aggregator of shops reviews (\textbf{YELP}); Amazon Music sample of songs reviews (\textbf{AMAZON}). The characteristics of the 3 data sets are summarised in Table \ref{tab:summary}.
\begin{table}[th]
\centering
\resizebox{0.5\linewidth}{!}{
\begin{tabular}{ccccc}
\toprule
 \textbf{data set} & \textbf{Interactions} &  \textbf{Users} &  \textbf{Items} &  \textbf{Sparsity} \\
\midrule
AMAZON &        64,706 &   5,541 &   3,568 &    0.0135 \\
ML-100K &       100,000 &    943 &   1,682 &    0.0630 \\
YELP &       267,211 &  16,442 &  11,128 &    0.0015 \\
\bottomrule
\end{tabular}
} 
\caption{Summary of data used for the experiments}
\label{tab:summary}
\end{table}

\noindent \textbf{Algorithms.} We test \texttt{FedFNN} with state-of-the-art algorithms and a baseline: $(i)$ \texttt{FedFast}, the current state-of-the-art, in terms of speed, to the best of our knowledge. It predicts the subordinates through a clustering based approach (\cite{fedfast});
$(ii)$\texttt{FedAvg}, baseline on which \texttt{FedFNN} and \texttt{FedFast} have been designed. It distributes local updates (both item and client embeddings) through averaging, and does not predict embeddings of the subordinate clients (\cite{mcmahan2017communication});
$(iii)$\texttt{WCU}, \emph{Without Client Updates} (WCU) algorithm only updates the embeddings of the sampled clients and the items that have been interacted with; subordinate embeddings are not updated. We include this option to emphasise the differences in performances with \texttt{FedFast}, which the only difference is the inclusion of the clustering method for predicting the unsampled clients; $(iv)$ \texttt{GMF}, the centralised version of the Neural Matrix Factorisation algorithm. It is included to establish the upper-bound of the FL-based solutions \cite{he2017neural}; $(v)$ \texttt{FedFNN-MLP}, \texttt{FedFNN} implementation using a multi-layer perceptron (MLP) to predict the unsampled clients; $(vi)$ \texttt{FedFNN-MLP-CV}, the architecture is the same of FedFNN-MLP, but this time the hyperparameters of \texttt{FedFNN} (dropout, number of hidden neurons, number of layers) are optimised running a grid-search in a 5-fold cross validation, at each training round.


For the item embedding updates it is important that an optimal strategy for redistributing the new item updates $(v^{t+1}[k])$ through a weighted average is found. For this reason we test different item embedding updates (\textbf{W0, W1} and \textbf{W2}).

\subsection{Performance and Accuracy (RQ1, RQ3)} 

\paragraph{Accuracy and Convergence Comparison.}

Here we compare different \texttt{FedFNN} implementations with the algorithms and data sets introduced above. In Figure \ref{fig:comparison} we show the results generated after both 100 and 500 iterations, where the first two columns of plots show the first 100 iterations (HR@10 and nDCG@10), while the second and third columns show the whole training rounds. 
In order to generate a fair comparison: $(i)$ for each data set, the learning rate of the local updates is the same for all the algorithms; $(ii)$ W1 is the item embedding strategy used for \texttt{FedFast}, \texttt{FedFNN} and \texttt{WCU} implementations; $(iii)$ the sample size of the local updates is fixed to $10\%$ for all the experiments.

Overall, \texttt{FedFNN-MLP} consistently outperforms the other algorithms on all 3 data sets, both in terms of speed of convergence (the number of iterations needed to reach stable level of convergence), and final accuracy (the final NDCG and HR). In our experiments we focus on the first 100 iterations of training to check the speed of convergence during this early stage: with ML-100K data, \texttt{FedFNN-MLP} is 2x faster than \texttt{FedFast}, for YELP data it is 5x faster, and for AMAZON it is even 8x faster than \texttt{FedFast}. Specifically, \texttt{FedFNN-MLP} needs only 12 iterations to reach \texttt{FedFast} accuracy at the 100th iteration. 
On the other hand, \texttt{FedFast} is better than \texttt{FedAvg} in terms of performance and accuracy, but presents similar performances of \texttt{WCU}, which is characterised by the absence of clustering techniques. This means that the clustering approach used to predict the clients' embeddings appears to have little effect on the performance of the algorithm. The introduction of the cross validation for \texttt{FedFNN}, helps to generate robust predictions (\texttt{FedFNN-MLP-CV}), presenting an increase in performance in the early stage of training, for all the experiments. In the case of MovieLens data set, \texttt{FedFNN-MLP-CV} is also slightly improving the overall performance of the standard \texttt{FedFNN-MLP}.


\paragraph{Grouped Preferences.} In order to analyse the performance of \texttt{FedFNN} in presence of clustered preferences we design a synthetic data generator which allows to input the number of groups of users presenting same preferences when generating the user-item interactions. 
We can first indicate the desired number of users (N) and items (M), then a mapping function $\ell : D \rightarrow \{D_1, D_2, ... D_G\}$, which maps each user-item interaction to a different subset $D_j$, where G is the total number of groups in input. 
For sake of simplicity, group of users do not overlap in terms of item preferences an all the groups have the same relative size. 
To generate more realistic data, we include parameters to model sparsity and popularity as a power-law distribution, which is generated through a \emph{Beta(a,b)} random variable. 
First, we use the Beta (with parameters fixed to $a = 1, b=3$) for generating the deviation from the average $s$ for all users, then, we assign to each user $u$ the final level of sparsity $s_u = s + s(s_u - s_M)$, where $s_M$ is the median of the sparsity vector generated through the sampling. 
The \emph{Beta} parameters are chosen to lead to a level of sparsity similar to the one in the real data set introduced in Table 1.
For each group of items $D_j$, we can define a different popularity bias vector, $\mathbf{p}_j$ which for each item $i$ in the group $j$, maps a coefficient $p_j(i) \in [0,1]$ representing the chances of being selected. In this way, when generating user preferences, first the number of interactions is computed ($n_u = \ceil{M\times s_u}$) and then $n_u$ items are sampled following $\mathbf{p}_j$.
The detailed process follows: $(1)$ we initialise the number of users $N$, the number of items $M$, the sparsity value $s$ and the number of groups G. Initialise the sparsity sample and generate the sparsity value for each user ($s_u$); $(2)$ for each group we generate the popularity vector $\mathbf{p}_j$, sampling from the beta distribution; $(3)$ for each user we then sample $n_u$ items following the probability:
$$
p(u,i) = \frac{h(u,i) p_i }{\sum_{j \in I} h(u,j) p_j}
$$
where $h(u,i) = \eta$ if $u$ and $i$ are mapped through $\ell(.)$ to the same group, otherwise $h(u,i) = (1- \eta)$. This parameter introduces a certain level of randomness, to have more realistic clusters distribution. In our experiments, we fix it to $\eta = 0.9$. 

We test \texttt{FedFNN} and \texttt{FedFast} on 6 different data sets, which differ only by the number of input clusters, where $G\in \{ 5, 10, 50, 100, 500, 1000\}$. In this case, \texttt{FedFast} is trained given in input the exact number of clusters used to generate the data. Also the parameters used to generate sparsity and popularity bias are the same along all the data sets. The sample size and local learning rate for the training of the algorithm does not change for the experiments. The results are presented in the Figure\ref{fig:clusters}, where the metric used is the difference in performance between \texttt{FedFNN} and \texttt{FedFast}, i.e. $\Delta_{HR}$ and $\Delta_{nDCG}$, positive if \texttt{FedFNN} presents better accuracy, negative if \texttt{FedFast} is more accurate. This value is measured at three different moments along the training: at early stage (after 100 iters); in the middle (after 300 iterations) and at the end of the training (after 500 iterations). It is possible to highlight how \texttt{FedFNN} is consistently better in terms of accuracy, looking at the values of $\Delta_{HR}$ and $\Delta_{nDCG}$. The improvement of \texttt{FedFNN} over \texttt{FedFast} is most noticeable in the early stages of learning (at 100 iterations), but even at later stages (after 300 and 500 iterations) it is consistently better in terms of accuracy. Since through the local updates, the two algorithms can only access to a portion of the whole data set, the clustering approach (\texttt{FedFast}) may struggle to sample necessary information to properly assign the clients to the right clusters.


\begin{minipage}{\textwidth}
\begin{minipage}[b]{0.6\textwidth}
\centering
\resizebox{0.9\linewidth}{!}{
\begin{tabular}{l|cc|cc|cc}
\toprule
\multicolumn{7}{c}{ML-100k}\\
\midrule
\multirow{2}{*}{Model} &  \multicolumn{2}{c|}{100 iter} &  \multicolumn{2}{c|}{300 iter} & \multicolumn{2}{c}{500 iter} \\
{} &  HR@10  &  nDCG@10  &  HR@10  &  nDCG@10 & HR@10  &  nDCG@10 \\
\midrule

W0+WCU     &  0.4740 &    0.2636 &  0.6352 &    0.3859 &  0.7137 &    0.4331 \\
W0+FedFast &  0.4995 &    0.2826 &  0.6691 &    0.4060 &  0.7158 &    0.4299 \\
W0+FedFNN &  \textbf{0.5801} &    \textbf{0.3421} &  \textbf{0.6893} &    \textbf{0.4180} &  0.7063 &    \textbf{0.4365} \\

\midrule

W1+WCU      &  0.5981 &    0.3487 &  0.7444 &    0.4445 &  0.7678 &    0.4666 \\
W1+FedFast  &  0.5790 &    0.3255 &  0.7264 &    0.4435 &  0.7614 &    0.4620 \\
W1+FedFNN  &  \textbf{0.7084} &    \textbf{0.4281} &  \textbf{0.7667} &    \textbf{0.4755} &  \textbf{0.7890} &    \textbf{0.4942} \\

\midrule
W2+WCU      &  0.5451 &    0.3050 &  0.7137 &    0.4262 &  0.7561 &    0.4648 \\
W2+FedFast  &  0.5323 &    0.2904 &  0.7010 &    0.4148 &  0.7529 &    0.4551 \\
W2+FedFNN  &  \textbf{0.6225} &    \textbf{0.3616} &  \textbf{0.7222} &    \textbf{0.4425} &  \textbf{0.7614} &    \textbf{0.4768} \\
\midrule
\multicolumn{7}{c}{Yelp}\\
\midrule
W0+WCU     &  0.1999 &    0.0916 &  0.2013 &    0.0921 &  0.2048 &    0.0952 \\
W0+FedFast &  0.2030 &    0.0937 &  0.2101 &    0.0977 &  0.2163 &    0.1021 \\
W0+FedFNN  &  \textbf{0.2147} &    \textbf{0.0986} &  0.2094 &    0.0975 &  \textbf{0.2241} &    \textbf{0.1058} \\
\midrule
W1+WCU      &  0.5042 &    0.2918 &  0.5768 &    0.3404 &  0.5880 &    0.3513 \\
W1+FedFast  &  0.5017 &    0.2895 &  0.5774 &    0.3403 &  0.5876 &    0.3491 \\
W1+FedFNN  &  \textbf{0.5910} &    \textbf{0.3488} &  \textbf{0.6123} &    \textbf{0.3703} &  \textbf{0.6342} &    \textbf{0.3902} \\

\midrule
W2+WCU      &  0.2036 &    0.0927 &  0.2890 &    0.1506 &  0.3963 &    0.2319 \\
W2+FedFast  &  0.2053 &    0.0937 &  0.2729 &    0.1373 &  0.3895 &    0.2262 \\
W2+FedFNN   &  \textbf{0.3099} &    \textbf{0.1636} &  \textbf{0.4082} &  \textbf{0.2381} &  \textbf{0.4385} &  \textbf{0.2606} \\

\bottomrule
\end{tabular}

}
    \label{table:ablation}
\captionof{table}{Summary of the ablation study conducted to analyse the impact of user and item embedding updates. In bold, the model presenting the best performances by different item embedding updates.}

\end{minipage}
\begin{minipage}[b]{0.5\textwidth}
    \begin{tabular}{c}
        \hspace{-3mm}\includegraphics[width=.7\linewidth]{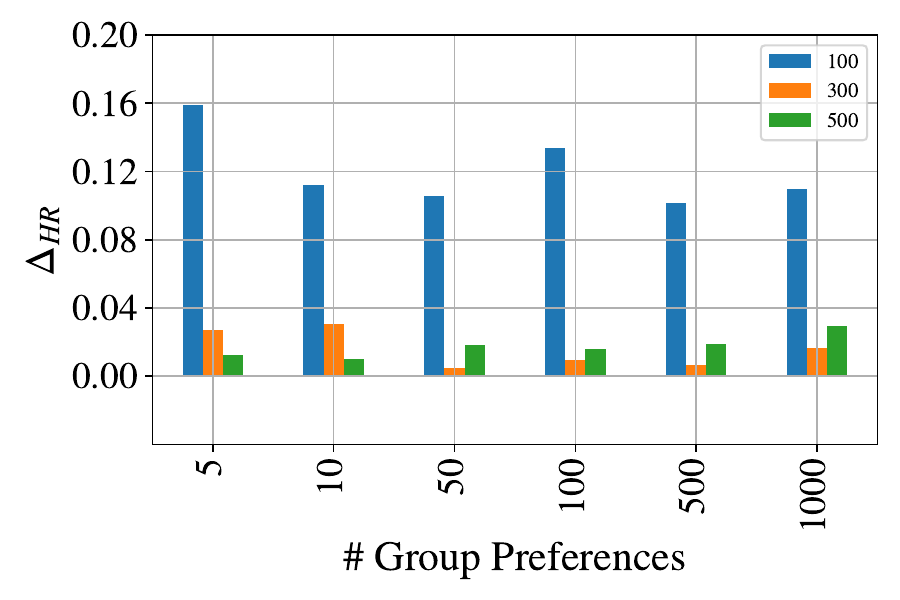}\\
        \hspace{-2mm}\includegraphics[width=.7\linewidth]{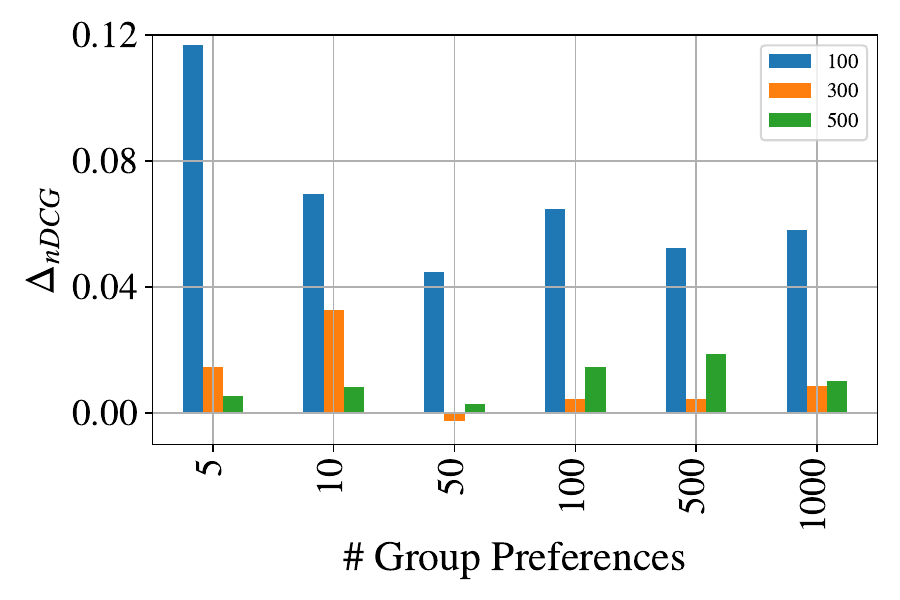} 
    \end{tabular}

\captionof{figure}{Comparsion of FedFast and \texttt{FedFNN} performances when users present different grouped preferences.}
    \label{fig:clusters}
    
\end{minipage}
\end{minipage}

\subsection{Ablation Study (RQ2)}
In order to study the contribution of the algorithm proposed, we perform an ablation study to emphasise the importance of using different strategies for both client and item embedding updates. For this reason we test three different strategies of subordinates predictions (\texttt{FedFNN}, \texttt{FedFast} and \texttt{WCU}) and three different strategies of item embedding updates (W0, W1 and W2). The experiments are performed following the same configurations used in Figure \ref{fig:comparison}. In Table \ref{table:ablation} we present the results for MovieLens and Yelp data sets, highlighting accuracy (HR@10 and nDCG @10) after 100, 300 and 500 iterations. About item embedding updates, for both data sets, \textbf{W1} presents better performances, in terms of speed and accuracy. Moreover, W2 presents better performances when compared to W0, which is consistently the slower one, no matter the choice of client predictions strategy. Among the client updates, \texttt{FedFNN} is optimal in terms of accuracy and speed of convergence, for both data sets, at all stages of training. \texttt{FedFast} and \texttt{WCU} demonstrate similar performance. 


\subsection{Poor availability and Sampling (RQ4)}

 We explore two different scenarios which are of practical significance in a FL setting: $(i)$ limited availability of clients, where along the training a group of client is consistently less available to be sampled (e.g. clients penalised when sampling the subset for the local updates); $(ii)$ a smaller fraction of clients are sampled to participate in training.

\paragraph{Availability.} Not all clients will have the same levels of availability during training, and those who participate less in training will have a diminished performing model. It is for this reason we decided to test how clients who participate less respond to the \texttt{FedAvg}, \texttt{FedFast} and \texttt{FedFNN} algorithms. Using ML-100K data set, we split (uniformly at random) the clients in two blocks of same size. One block is fully available along the training (\textbf{normal availability}), while all the clients belonging to the other have 25\% chance of being available at each iteration for being sampled and run the local update (\textbf{poor availability}). Partitioning the clients this way means that the clients assigned to the poor availability group are consistently sampled and locally updated less frequently than the other group. We test and monitor the performances of three different models in this case (\texttt{FedFNN}, \texttt{FedFast} and \texttt{FedAvg}). In Figure 4b we report on the left (Fig. 4a) the performance for the portion of users consistently available, while on the right (Fig. 4b) the performance of the sample \emph{poorly available}. \texttt{FedFNN} shows better performances for the both groups of clients. In particular, the gap in terms of accuracy is larger for the poor availability clients, especially at the early stages of the training. For both groups of clients, \texttt{FedFast} and \texttt{FedFNN} stabilize around the same level of accuracy, while \texttt{FedAvg} fails to reach the same level of accuracy for the poor availability group.

\noindent\textit{Sampling.} In this experiment we explore how different sampling rates affect the convergence of \texttt{FedFNN}, \texttt{FedFast} and \texttt{FedAvg}. Note that the sampling rate is the fraction of clients which participate in the training at each iteration. We choose 3 different sampling sizes, $\{1\%, 5\%, 20\% \}$. In Figure 4a we present the experiments. For the smallest sample size (Fig. 5a) \texttt{FedFast} and \texttt{FedAvg} present high level of variability along all the training rounds, while \texttt{FedFNN} is stable, converging faster than \texttt{FedFast} and \texttt{FedAvg} to higher level of accuracy. For the other sample sizes (Fig. 5b and Fig. 5c) the differences in performance between \texttt{FedFNN} and the other algorithms still exists but it is increasingly diminished.
\begin{figure}[t]
    \centering
    \begin{minipage}[b]{.45\textwidth}
    \centering
    \includegraphics[width=0.8\textwidth]{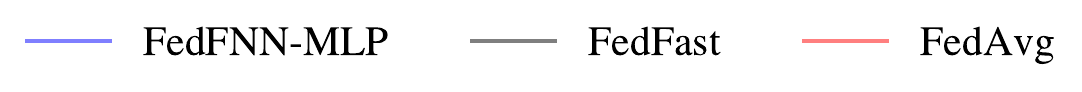}\\
    \includegraphics[width=0.8\linewidth]{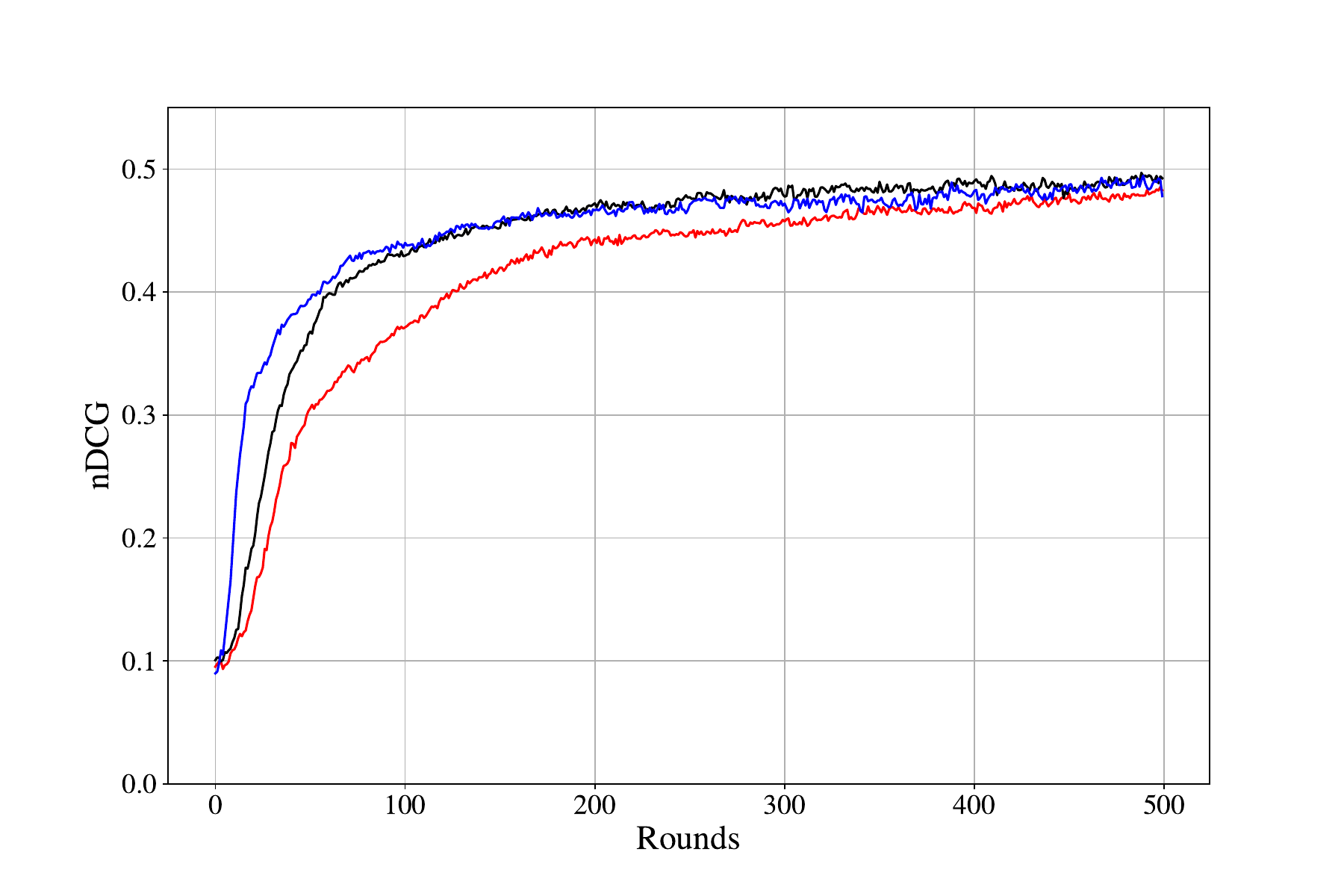} \\
    \includegraphics[width=0.8\linewidth]{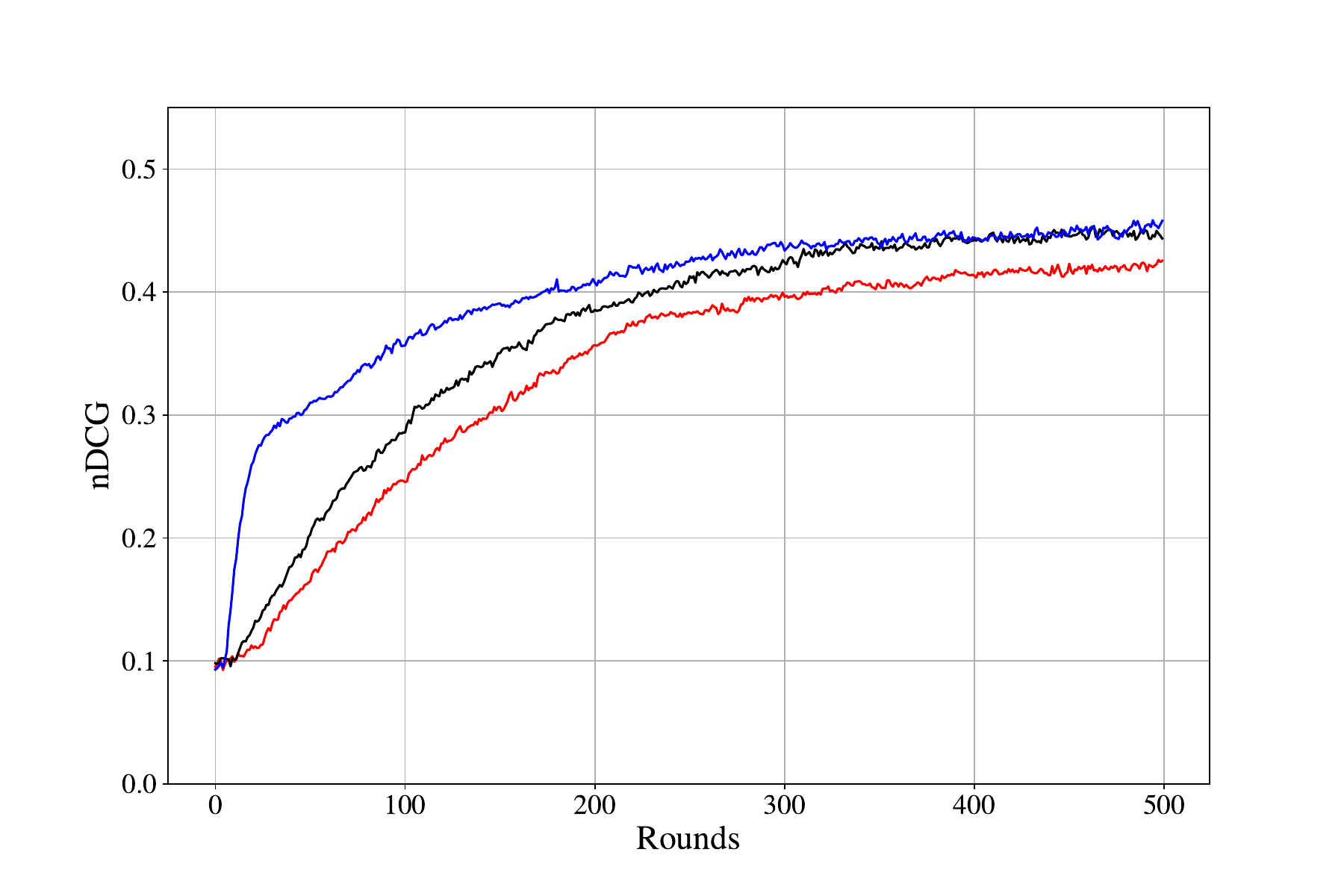}\\
    (a)

    \label{fig:pooravail}
    \end{minipage}
    \begin{minipage}[b]{.5\textwidth}
    \centering
    \includegraphics[width=0.7\linewidth]{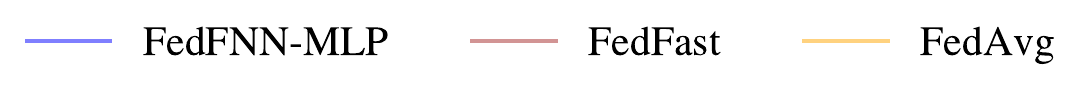}\\
    \includegraphics[width=0.5\linewidth]{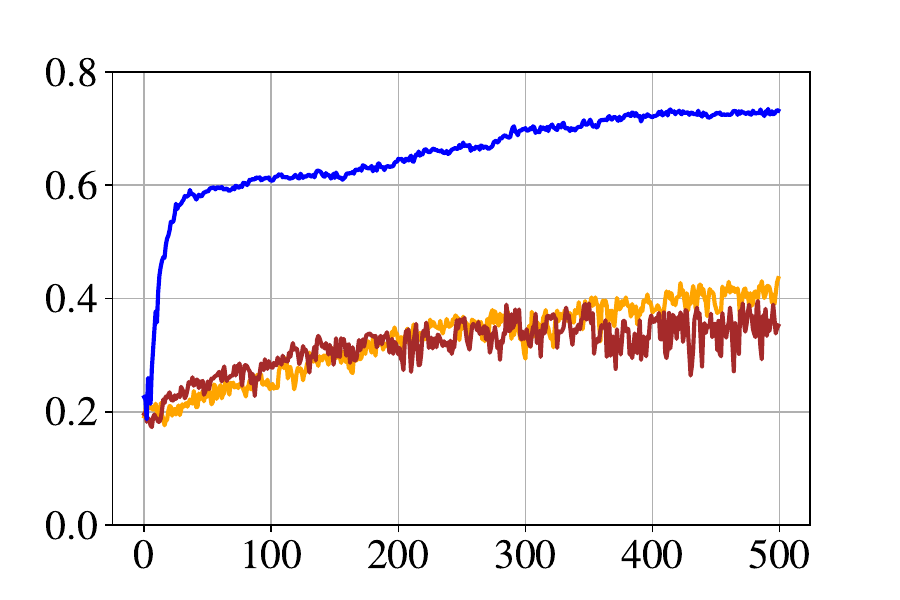}\\
    \includegraphics[width=0.5\linewidth]{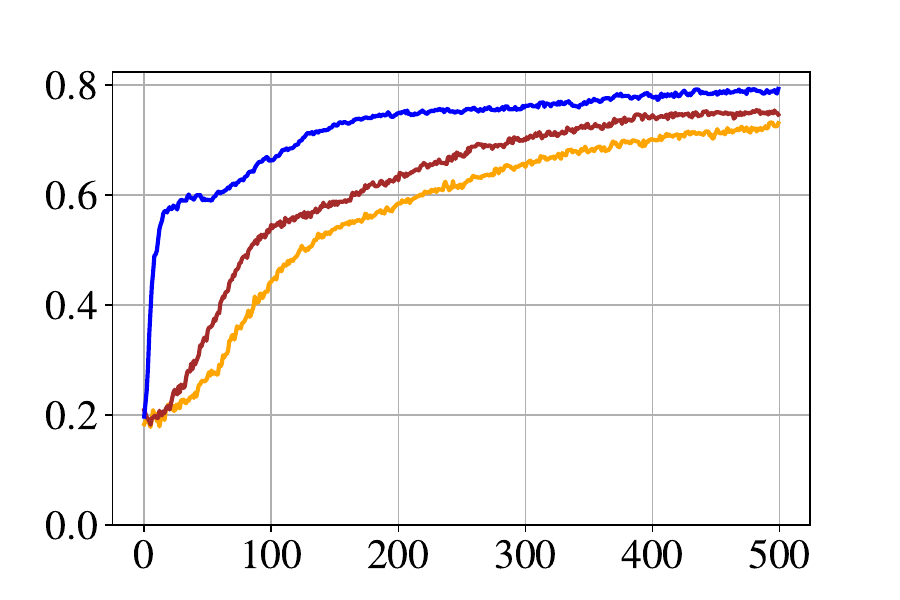}\\
    \includegraphics[width=0.5\linewidth]{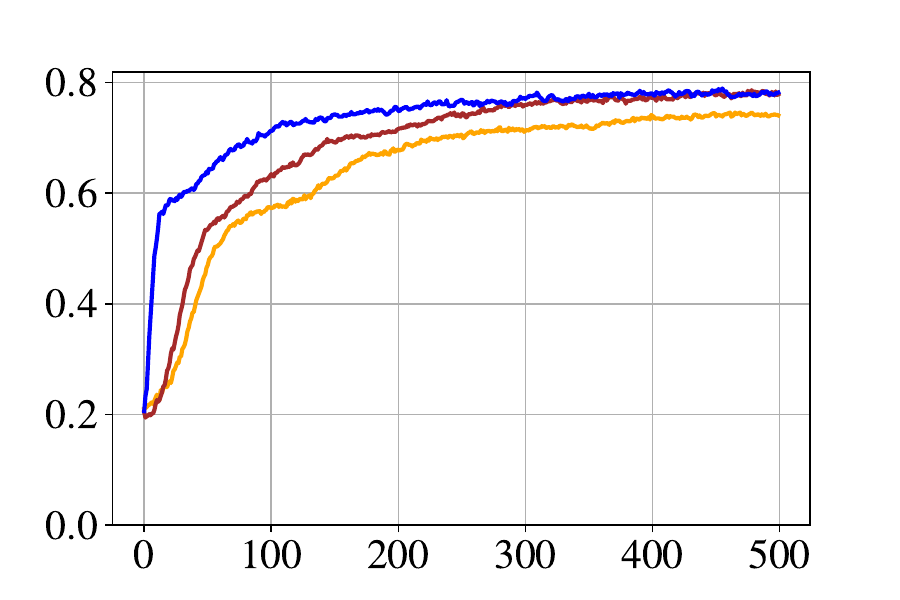}\\
    (b)

    \label{fig:sample}    
    \end{minipage}
\caption{Simulations for: different configurations of poor availability (left) and 3 distinct sample sizes (right). The models are tested on FedFNN, FedFast and FedAvg. }
\end{figure}



\section{Discussions and Conclusions}
\label{sec:disc}
This paper presents \texttt{FedFNN}, a new framework in Federated Learning, which speeds up the training of machine learning models, with a focus on the recommender systems.
\texttt{FedFNN} framework predicts the weight updates for clients not available during training, increasing the performance of the model, especially at the early stage of the process, proving to be not only faster, but with higher accuracy in the same amount of training rounds.
Moreover, \texttt{FedFNN} is shown to perform optimally in conditions where the sample size is relatively small or when some clients present poor availability.
We also demonstrate empirically that training \texttt{FedFNN} over input data presenting a high number of clustered preferences (e.g. distinct group of clients with distinguishable preferences) is still beneficial when compared with existing alternatives. 
Given its novelty, there are several limitations of our model that can be address in the future work.
Firstly, it is trained with offline data (ratings or preferences existing before the start of the training), which does not include any online interactions. 
We introduce \texttt{FedFNN} in the context of recommendation algorithms based on matrix factorisation.
Although matrix factorisation remains a popular approach in the literature, a new family of algorithms based on deep neural networks recently gaining traction. 
In the future, we will consider the adaptation of \texttt{FedFNN} to these approaches, such as graph neural networks (GNN) or Deep \& Cross Networks  (DCN) \cite{wu2021fedgnn, wang2021dcn}. \texttt{FedFNN} has a significant impact on the first iterations of the model.

\bibliographystyle{splncs04}
\bibliography{bibby}


\end{document}